\documentclass[conference]{IEEEtran}
\IEEEoverridecommandlockouts

\usepackage{amsmath,amssymb,mathtools} % mathtools must come with amsmath
\usepackage{graphicx}
\usepackage{booktabs}
\usepackage{multirow}
\usepackage{subcaption} % but do NOT load 'caption'
\usepackage{tabularx} % Add in preamble
\usepackage{booktabs} % For better table rules

\usepackage[ruled,vlined]{algorithm2e}
\DontPrintSemicolon

\SetKwFunction{Cost}{Cost}
\SetKwProg{Fn}{Function}{}{}

\usepackage{enumitem}
\usepackage{siunitx} % v2/v3 both fine for this doc
\usepackage{xspace}
\usepackage{xcolor}

\usepackage[hyphens]{url}
\usepackage{hyperref}
\usepackage[nameinlink]{cleveref}

\usepackage{tikz}
\usepackage{pgfplots}
\pgfplotsset{compat=1.18} % recommended

\newcommand{\system}{\textsc{AdaScale}\xspace}

\hypersetup{
  colorlinks=true,
  linkcolor=black,
  citecolor=black,
  urlcolor=black
}

\begin{document}

%%%%%%%%%%%---SETME-----%%%%%%%%%%%%%
%%%%%%%%%%%%%%%%%%%%%%%%%%%%%%%%%%%%

\pagenumbering{arabic}

\title{\system: An Adaptive Scaling and Placement Framework for Microservices Under Dynamics
%under Dynamics
% AdaScale: Two-Timescale Joint Scaling and Placement for Cloud--Edge Microservices under Dynamic Call-Graph Mixes
% AdaScale: SLO-Driven Joint Autoscaling and Network-Aware Placement for Dynamic Microservices

}
% \author{\normalsize{Anonymous Authors}}

\author{Ming Chen, Muhammed Tawfiqul Islam, Maria Rodriguez Read, Rajkumar Buyya,~\IEEEmembership{Fellow,~IEEE}
        % <-this % stops a space
% \thanks{Manuscript received July xx, 2025; revised August xx, 20xx.}
    % \thanks{Manuscript submitted July, 2025.}
    \thanks{Ming Chen, Muhammed Tawfiqul Islam, Maria Rodriguez Read, and Rajkumar Buyya are with the Quantum Cloud Computing and Distributed Systems (qCLOUDS) Laboratory, School of Computing and Information Systems, The University of Melbourne, Australia. 

    }}
%(e-mail: mingc4@student.unimelb.edu.au; {tawfiqul.islam, maria.read, rbuyya}@unimelb.edu.au)}

\maketitle
\thispagestyle{plain}
\pagestyle{plain}

% ...Abstract and sections...

\begin{abstract}
Microservice applications are increasingly deployed across cloud--edge environments, where heterogeneous nodes and time-varying inter-node delays amplify the impact of placement decisions. At the same time, these applications face non-stationary traffic, shifts in the mix of root request operations that exercise different call graphs, and heterogeneous communication modes that determine how network latency and queuing propagate to end-to-end (E2E) performance. Existing autoscalers and network-aware schedulers typically handle only a subset of these dynamics, leading to either compute bottlenecks or inflated cross-node latency and thus SLO violations.

We propose \system, an adaptive framework that jointly scales and places microservice replicas under such multi-dimensional dynamics. \system implements a Monitor--Analyzer--Planner--Executor (MAPE) loop that extracts per-edge and per-service demand from distributed traces and service-mesh metrics, identifies the most critical root operation under a mixed workload, computes SLO-aware replica targets, and then places replicas to minimize a demand-weighted latency objective given the current inter-node latency matrix. To react quickly to networking perturbations, \system triggers a reactive placement loop, while a steady-state autoscaling loop handles demand shifts.

We evaluate \system on a cloud--edge Kubernetes cluster using the DeathStarBench Social Network application with three root operations under varying load and workload mixes. Across scenarios, \system consistently meets SLO targets and improves both latency and throughput: compared with NetMARKS\_Scale, it achieves up to 1.56$\times$, 1.93$\times$, and 1.34$\times$ lower average response time (for compose-post, read-home-timeline, and read-user-timeline) and up to 2.16$\times$, 1.32$\times$, and 1.36$\times$ higher throughput, respectively.
\end{abstract}


\section{Introduction}

Microservice architecture is widely adopted for large-scale online services, decomposing an application into loosely coupled services that can be developed and deployed independently~\cite{Parslo_SoCC21, SoCC2021_MS_luo}. Modern deployments increasingly span a cloud--edge continuum, where replicas run on heterogeneous nodes to satisfy latency, cost, and locality requirements. In such environments, end-to-end (E2E) quality depends not only on per-service compute provisioning but also on replica placement: time-varying inter-node delays can quickly turn previously acceptable placements into latency bottlenecks~\cite{NetMARKS_InfoCOM21, Net_ms_placement_TNSM,TraDE_TPDS}.

Managing microservices under dynamics is difficult because several variabilities interact. User demand is non-stationary and bursty, requiring timely replica provisioning. Meanwhile, production applications usually serve multiple root operations (e.g., read vs.\ write paths) with different call graphs and often different E2E SLOs; shifts in the workload mix therefore change which services and edges dominate E2E latency. Dependencies also span heterogeneous communication modes (e.g., blocking RPC, message queues, and storage I/O), so identical network perturbations can manifest differently across edges. These effects are coupled: poor placement amplifies cross-node delay even with sufficient replicas, while under-provisioning creates compute bottlenecks even under favorable network conditions.

Existing orchestration mechanisms typically handle only part of this problem. Kubernetes Horizontal Pod Autoscaler (HPA)~\cite{K8s_HPA} scales replicas based on utilization signals, but is blind to network state and to which root operation is currently SLO-critical. Network-aware schedulers~\cite{NetMARKS_InfoCOM21, Net_ms_placement_TNSM} can improve placement using service-mesh telemetry, yet often assume fixed replica budgets and do not explicitly reason about changing request mixes. Conversely, call-graph and SLO-analysis techniques can localize bottlenecks, but are often not designed to drive coordinated scaling and placement decisions under time-varying network states. These gaps motivate a unified framework that continuously translates runtime call-graph behavior and network state into joint scaling and placement actions.

This paper proposes \system, an adaptive scaling and placement framework for distributed microservice applications under dynamics. \system follows a Monitor--Analyzer--Planner--Executor (MAPE) loop that collects distributed traces, service metrics, and inter-node latency measurements; infers per-edge and per-service demand summaries from observed call graphs; derives SLO-aware replica targets under the current workload mix; and places replicas to reduce demand-weighted cross-node latency under current network conditions. A key design principle is two-timescale adaptation: \system uses a steady-state autoscaling loop to handle demand and workload-mix shifts, and a reactive placement loop to promptly mitigate network perturbations that threaten E2E SLO compliance.

We implement \system on Kubernetes with a service-mesh and tracing stack, and evaluate it on a cloud--edge cluster using the DeathStarBench Social Network benchmark~\cite{deathStarBench_ASPLOS19}, which exposes three root operations with distinct call graphs. Across a range of request rates and workload mixes, \system consistently meets SLO targets and improves both latency and throughput compared with Kubernetes HPA and NetMARKS\_Scale. This paper mainly makes the following contributions:
\begin{itemize}
  \item We formulate joint scaling and placement as an optimization problem that balances per-service resource cost against demand-weighted network latency under E2E and per-service quantile SLO constraints, and decompose it into tractable scaling and placement subproblems for online control.
  \item We design and implement \system as a two-timescale MAPE controller with (i) an SLO- and demand-aware autoscaler that identifies the most critical root operation under mixed workloads and computes per-service replica targets, and (ii) a greedy, latency- and capacity-aware placer that maps replicas to nodes using the current inter-node latency matrix.
  \item We evaluate \system on a real Kubernetes-based cloud--edge testbed, showing up to 1.93$\times$ lower average response time and up to 2.16$\times$ higher throughput than existing methods while maintaining SLO compliance under dynamic conditions.
\end{itemize}

The rest of the paper is organized as follows: Section~\ref{sec:Background and challenges} presents background and challenges; Section~\ref{sec:system-model} introduces the system model and telemetry-based demand estimation; Section~\ref{sec:formulation} formulates the joint problem; Section~\ref{sec:architecture} details \system's design; Section~\ref{sec:performance evaluation} evaluates performance; Section~\ref{sec:realted work} reviews related work; and Section~\ref{sec:conclusion} concludes the paper.

\section{Background and Challenges}
\label{sec:Background and challenges}

\subsection{Background}
Modern cloud applications increasingly adopt microservice architectures, managing an application as a collection of containerized, loosely coupled, fine-grained services~\cite{deathStarBench_ASPLOS19, SoCC2021_MS_luo}. In cloud--edge continuums, services may run in centralized cloud data centers, on fog/edge nodes close to end users, or across both. A common hybrid pattern places latency-critical, user-facing services at the edge while keeping centralized or stateful components in the cloud, reducing user-perceived latency. However, end-to-end performance can drift over time due to both networking dynamics and workload dynamics (e.g., changing request intensity and call-graph mix).

\subsection{Challenges}\label{subsec:motivation}
\subsubsection{Networking Uncertainty}
Varied networking conditions are common in cloud--edge clusters and interconnected fog devices. Microservice applications allocate a higher proportion of processing time to the networking stack than monolithic applications~\cite{deathStarBench_ASPLOS19}, since requests frequently traverse RPC/REST interfaces across services. As inter-node latency and available bandwidth vary over time, cross-node communication can become a dominant contributor to E2E latency and throughput degradation, even when per-service compute provisioning is sufficient.

\begin{figure}[t]
    \centering
    \includegraphics[width=\linewidth]{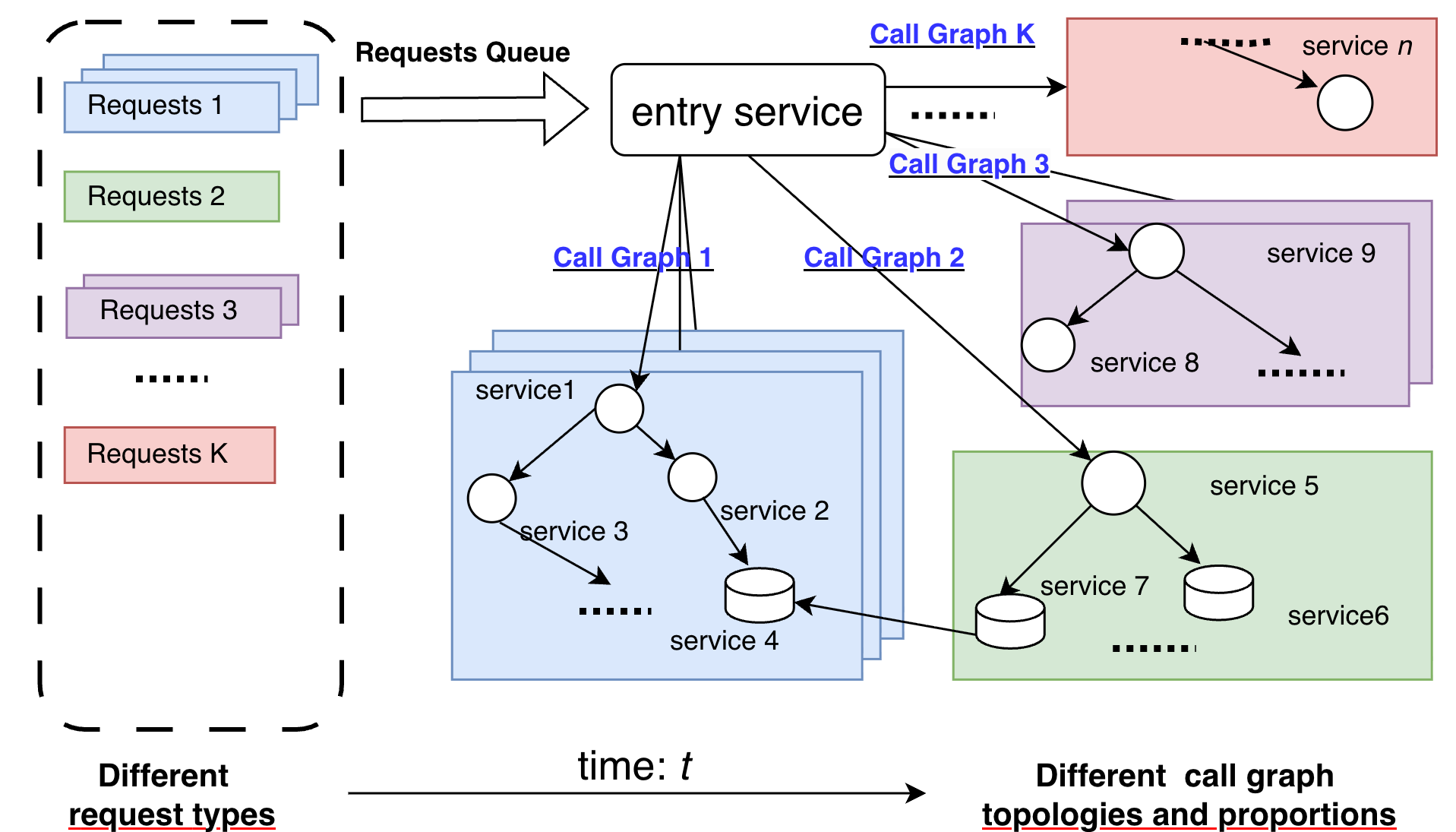}
    \caption{Root requests induce a time-varying mix of call graphs $\mathcal{G}=\{G_1,\dots,G_K\}$ with proportions $\boldsymbol{\pi}(t)$, where $\sum_{k=1}^K\pi_k(t)=1$.}
    \label{fig:dynamic_graph}
\end{figure}

\begin{figure}[t]
    \centering
    \includegraphics[width=\linewidth]{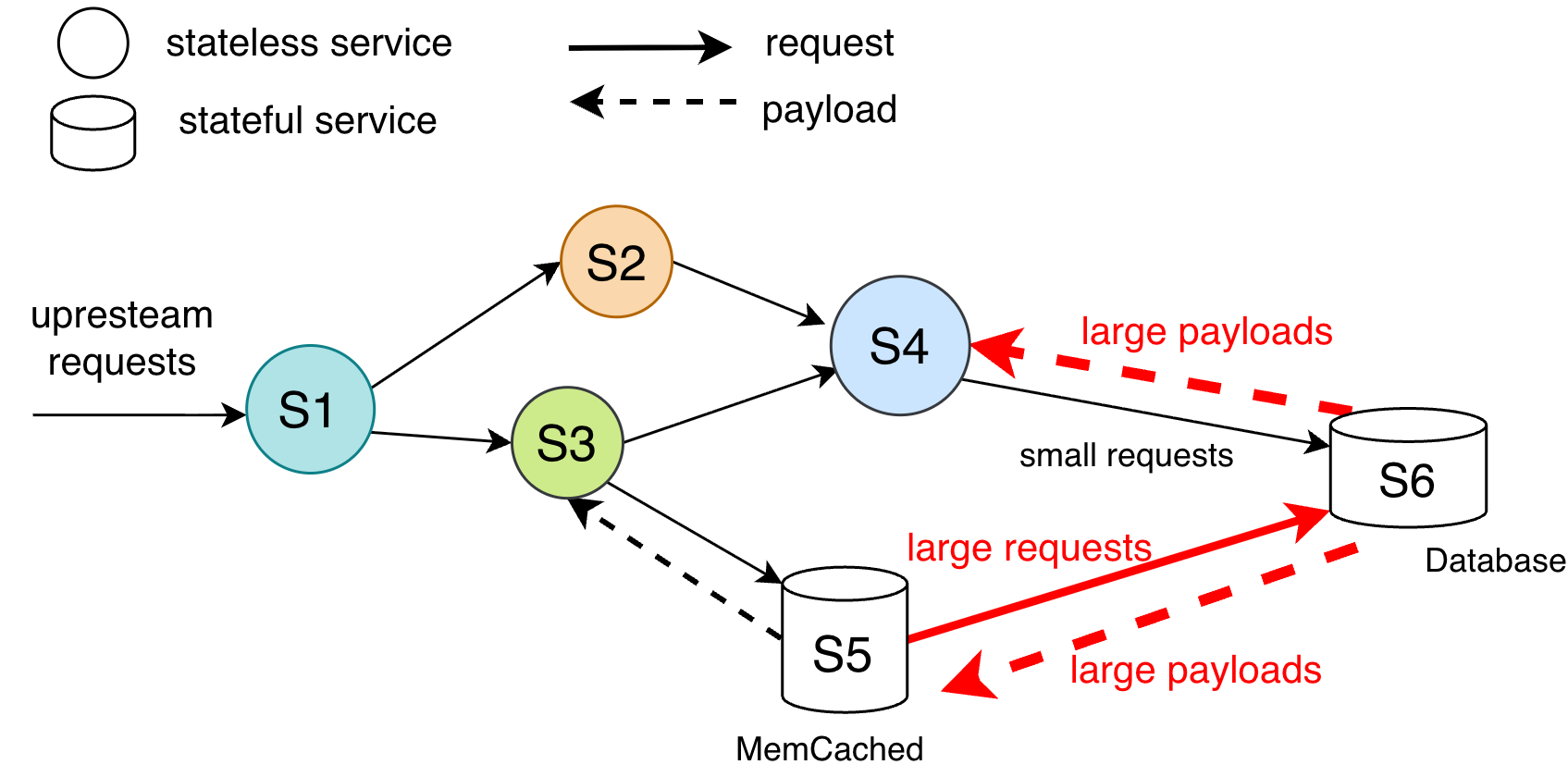}
    \caption{Call-graph edges exhibit heterogeneous communication modes and traffic patterns.}
    \label{fig:motivation3_traffic}
\end{figure}

\subsubsection{Dynamics of Call Graphs}
In a microservice dependency graph, if one service triggers another, we refer to them as \textit{UM} (upstream microservice) and \textit{DM} (downstream microservice), respectively. The entry service receives incoming \textit{root requests}, which are routed through downstream services to execute application logic. As shown in Figure~\ref{fig:dynamic_graph}, different root-request types exercise different call graphs, and the runtime workload is typically a time-varying mixture over these graphs.

This mixture is dynamic in both (i) \textit{topology} (which call paths are exercised) and (ii) \textit{proportions} (how frequently each path occurs). Shifts in request mix therefore change which services or service pairs become hot and which paths dominate E2E latency. Shared services that appear in multiple call graphs (e.g., stateful storage) further couple demand across operations, making scaling and placement decisions dependent on the current mix rather than on a single static graph.

\subsubsection{Communication Modes}
Production traces from Alibaba and Meta show that microservice pairs interact through heterogeneous communication modes (e.g., HTTP, gRPC, DB, MQ, and MC)~\cite{github_metaTraces, github_AlibabaTraces, SoCC2021_MS_luo, metaTraces_ATC23}. These modes induce different traffic shapes and sensitivities to load and network variation. In particular, DB/MC interactions are often heavy in traffic and can become bottlenecks under demand bursts or networking perturbations, degrading upstream E2E performance. Moreover, as depicted in Figure~\ref{fig:motivation3_traffic}, request/response sizes can be asymmetric (e.g., small requests but large payloads), so a placement/scaling policy should account for bidirectional traffic characteristics along each edge.

Overall, effective management in cloud--edge settings must jointly consider: (1) time-varying inter-node latency and bandwidth; (2) dynamic call-graph topology and proportions under mixed root requests; and (3) heterogeneous communication modes that shape traffic and bottlenecks. These coupled challenges motivate a systematic framework that estimates network and call-graph dynamics at runtime and drives coordinated scaling and placement decisions.

\section{System Model}
\label{sec:system-model}

We consider a \emph{microservice application} composed of stateless and stateful microservices
$\mathcal{S}=\{s_1,\dots,s_N\}$ deployed on a cluster of compute nodes $\mathcal{N}=\{n_1,\dots,n_M\}$. Each microservice $s\in\mathcal{S}$ has $x_s(t)\in\mathbb{N}$ replicas (e.g., Kubernetes Pods) at time $t$.
A single replica of $s$ consumes a per-replica resource vector $R_s=[c_s,\,m_s]$ (CPU shares and memory), while node $n$ offers capacity $C_n=[\bar c_n,\,\bar m_n]$. For network conditions of the cluster, we primarily consider the inter-node communication delays and assume these delays evolve randomly as time goes by.

\subsection{Replica placement matrix.}
In \system implementation, a service may spread its replicas across multiple nodes. We therefore represent a placement at time $t$ by a matrix \[
A(t)=\bigl[A_{s,n}(t)\bigr]_{s\in\mathcal{S},\,n\in\mathcal{N}},
\]
where $A_{s,n}(t)\in\mathbb{N}$ is the number of replicas of service $s$ hosted on node $n$.
The total replica count of $s$ is $x_s(t)=\sum_{n\in\mathcal{N}}A_{s,n}(t)$.
In the implementation, the data fields \texttt{assignments:\{node $\mapsto$ replica count\}} and \texttt{replicas} in the placement files encode $A_{s,n}(t)$ and $x_s(t)$, respectively.

%The concrete JSON encodings \texttt{placement.json} and \texttt{placement\_plan.json} represent, respectively, the current $A(t)$ and the planned $A'(t)$ at a decision epoch, as produced and
%consumed by AdaScale.\footnote{In the implementation, each service records its \texttt{assignments:\{node $\mapsto$ replica count\}} and \texttt{replicas} fields in these JSON files.}

\subsection{Workflows as mixtures of call graphs}
In a deployed microservice application, multiple root request types can coexist and thus trigger different call graphs. We refer to a root request as a request entering the application at its entry service (e.g., a front-end gateway).  
Incoming root requests arrive at a total rate $\lambda_{\text{root}}(t)$ and are partitioned into $K$ request classes (e.g., compose-post, read-home-timeline, read-user-timeline) with per-class rates $\lambda_k(t)$ and proportions
\[
\pi_k(t) \;=\; \frac{\lambda_k(t)}{\max\{\lambda_{\text{root}}(t),1\}}, \quad \sum_{k=1}^K \pi_k(t)=1.
\]
Each root-request class~$k$ triggers its own call graph $G_k=(V_k,E_k)$.
Thus, incoming root requests $\lambda_{\text{root}}(t)$ trigger a time-varying mixture over the microservice call graphs
$\mathcal{G}=\{G_1,\dots,G_K\}$ with proportions
$\boldsymbol{\pi}(t)=(\pi_1(t),\dots,\pi_K(t))$.
Each call graph $G_k=(V_k,E_k)$ has $V_k\subseteq\mathcal{S}$ and
directed edges $e=(u\!\to\!v)\in E_k$, where $u$ and $v$ denote
upstream and downstream microservices. Over a time window $W_t$, we compute the following edge-level statistics from tracing spans:

Each edge $e$ is annotated with an attribute
\begin{equation}
  \label{eq:edge-attr}
  \text{attr}(e)=\langle \text{mode}_e,\; r_e(t),\; \theta_e,\;
  w_e(t),\; b_e(t)\rangle,
\end{equation}
where:
\begin{itemize}[leftmargin=*]
  \item $\text{mode}_e\in\{\mathrm{HTTP},\mathrm{gRPC},\mathrm{MQ},\mathrm{DB},\mathrm{MC}\}$ captures the communication mode; different modes have distinct traffic tramit patterns.
  \item $r_e(t)$ is the estimated call rate (calls/second) on edge $e$;
  \item $\theta_e$ identifies the callee interface (e.g., HTTP path, gRPC
        method, DB collection/op);
  \item $w_e(t)$ is the average edge-local service time (milliseconds), which can be obtained from child span durations in raw metrics traces;
 % The term $w_e(t)$ is the mean child-span duration recorded for edge $e$ over $W_t$, and is used as a proxy for callee-side work per call.

  \item $b_e(t)$ is the byte rate (bytes/s) on $e$, which obtained from
        service-mesh traffic counters.
\end{itemize}

\subsection{Edge statistics from traces}
\label{subsec:edge-stats}

From each metrics span (e.g., Jaeger traces), we estimate, over a time window $W_t$, the following edge statistics:
\begin{equation}
\label{eq:edge_stats}
\begin{split}
p_e(t)
&\triangleq
\frac{\# \text{root requests that traverse }e}{\max\{\#\text{root
requests},1\}},\\[-0.25em]
r_e^{\text{per-req}}(t)
&\triangleq
\frac{\# \text{occurrences of }e}{\max\{\# \text{requests that
traverse }e, 1\}},\\[-0.25em]
w_e(t)
&\triangleq
\text{mean child-span duration for edge }e.
\end{split}
\end{equation}
A root request is a request entering the application at its entry service. The quantity $p_e(t)$ is the probability that a root request exercises edge $e$, $r_e^{\text{per-req}}(t)$ captures repeated invocations \emph{conditional} on traversing $e$, and $p_e(t)\cdot r_e^{\text{per-req}}(t)$ is the expected number of occurrences of $e$ per root request.  
For the trace sampling strategies, we use probabilistic sampling, which means sampling traces based on a defined probability (e.g., $0.1$ for 10\% sampling). If the root request rate is $\lambda_{\text{root}}(t)$ and the tracing sampling probability is $\rho_{\text{sample}}$,  then the call rate on edge $e$ is estimated as
\[
r_e(t)
\approx
\lambda_{\text{root}}(t)
\cdot
\frac{p_e(t)\,r_e^{\text{per-req}}(t)}{\rho_{\text{sample}}}.
\]

Here $\lambda_{\text{root}}(t)$ denotes the true root request rate (requests/s) for the application. We assume sampling is approximately uniform over requests, so rescaling by $1/\rho_{\text{sample}}$ debiases the per-edge counts.

\subsection{Service-level demand}
\label{subsec:service-demand}
Under sustained workloads on the deployed microservice application, we can obtain per-service demand metrics by aggregating call graph edge statistics.
For each service $s\in\mathcal{S}$ we define:
\begin{align*}
R_s^{\text{in}}(t) &= \sum_{e=(u,s)} r_e(t),
&
R_s^{\text{out}}(t) &= \sum_{e=(s,v)} r_e(t),\\
W_s^{\text{in}}(t) &=
\frac{\sum_{e=(u,s)} r_e(t)\,w_e(t)}{\max\{R_s^{\text{in}}(t),1\}},
&
B_s^{\text{in}}(t) &= \sum_{e=(u,s)} b_e(t),
\end{align*}
%\textcolor{red}{$W_s^{\text{in}}(t)$ is the average incoming work per call (ms)}
where $R_s^{\text{in}}(t)$ and $R_s^{\text{out}}(t)$ are the total incoming and outgoing call rates of service $s$, \textcolor{black}{$W_s^{\text{in}}(t)$ is the average incoming work per call (ms)}, and $B_s^{\text{in}}(t)$ is the incoming byte rate which can be used as an importance weight when prioritizing services under stateful backends or high network overhead. In practice of \system implementation, these quantities are produced by the \texttt{service\_edge\_demand} module as the \texttt{services\_demand} table, while the per-edge quantities $r_e(t)$, $w_e(t)$, $b_e(t)$ appear in the \texttt{edges\_demand} table.

An approximate CPU demand (CPU-seconds per second) for service $s$ is then \[
\text{CPU\_demand}_s(t)
\approx
R_s^{\text{in}}(t)\cdot\frac{W_s^{\text{in}}(t)}{1000}
% \quad\text{[CPU-seconds per second],}
\]
which directly benefits the decisions of resource provisioning for replicas of  service $s$. Here $w_e(t)$ and $W_s^{\text{in}}(t)$ are measured in milliseconds, so $R_s^{\text{in}}(t)\cdot W_s^{\text{in}}(t)/1000$ has units of CPU-seconds per second, i.e., an effective fraction of one CPU core under the simplifying assumption that $w_e(t)$ is dominated by CPU service time.

\subsection{Networking state (latency-only)}
\label{subsec:network-state}

For each node pair $(i,j)\in\mathcal{N}\times\mathcal{N}$, a
set of distributed agents continuously measures inter-node latency $L_{i,j}(t)$ using ICMP probes at a
configurable interval. An inter-node latency matrix maintains these cross-node delay
measurements:
\[
L(t)=\{L_{i,j}(t)\mid i,j\in\mathcal{N}\}.
\]
\system intentionally does not construct or use an explicit bandwidth prior to the current stage for three reasons: 
(i) most latency-sensitive microservice applications in our target setting are bottlenecked by CPU and queueing rather than raw network bandwidth; (ii) considering varying bandwidth across cluster nodes would significantly complicate the scaling decision process; (iii) evolving cross-node latencies across the cluster nodes is enough to emulate a dynamic computing environment. Considering the stateful services existing in the deployed microservice application,  byte counters are used through $b_e(t)$ and $B_s^{\text{in}}(t)$
to weight the contributions of the cross-node latencies, while the control objective is driven
entirely by \emph{latency} and \emph{resource} metrics.

Given a placement matrix $A(t)$, the expected network latency for a
single invocation on edge $e=(u,v)$ between service $u$ and service $v$ is approximated as
\begin{equation}
\label{eq:edge-latency-multi}
\bar L_e\bigl(A(t),L(t)\bigr)
\approx
\sum_{i,j\in\mathcal{N}}
\frac{A_{u,i}(t)}{x_u(t)}
\cdot
\frac{A_{v,j}(t)}{x_v(t)}
\cdot
L_{i,j}(t).
\end{equation}

Intuitively, $\bar L_e(A,L)$ is the expected network delay per invocation on edge $e=(u,v)$ if caller and callee replicas are chosen uniformly at random among the replicas of $u$ and $v$ under placement $A$.

\subsection{Latency composition}
\label{sec:latency-model}

For a root request that traverses call graph $G_k$, its end-to-end latency can be
estimated by aggregating contributions along invocation paths:
\begin{equation}
\label{eq:latency}
\begin{aligned}
Y_k(t)
&\approx
\sum_{(u,v)\in \mathcal{P}_k}
\Bigl[
  \bar L_{(u,v)}\bigl(A(t),L(t)\bigr)
\\
&\qquad
  {}+ \phi_{\text{mode}_e}\bigl(x_u(t),x_v(t),r_e(t)\bigr)
\Bigr].
\end{aligned}
\end{equation}
where $\mathcal{P}_k$ denotes the multiset of edges on the request's execution path within $G_k$, $\bar L_{(u,v)}$ is the expected replica-averaged inter-node delay in Equation~\eqref{eq:edge-latency-multi}, and $\phi_{\text{mode}_e}(\cdot)$ captures callee-side queueing and protocol semantics for edge $e$ (e.g., blocking RPC delay, MQ enqueue/dequeue delay, storage I/O) as a function of replica counts and call rate. Note that $\phi_{\text{mode}_e}(\cdot)$ depends on the call rate $r_e(t)$ (through utilization and queueing), whereas $w_e(t)$ is used only in the demand estimates of Section~\ref{subsec:service-demand}.

In practice, \system does not attempt to solve Equation~\eqref{eq:latency} exactly. Instead, it uses \emph{span-level decomposition}, which attributes per-request latency contributions to individual services using traces. The approximations, together with the demand matrices described above, are sufficient to drive the optimization in real time.

%and (ii) \emph{service-level response models}, which fit the dependence of per-service latency on load and replica counts. These approximations, together with the demand matrices described above, are sufficient to drive the optimization in real time.

\subsection{Correlations with runtime dynamics}
\label{subsec:correlations}

The end-to-end latency distribution is shaped by several coupled
factors:
\begin{itemize}[leftmargin=*]
  \item Replica counts $x(t)=\{x_s(t)\}$ determine per-service queueing
        and contention, and thus affect the $\phi_{\text{mode}_e}$ terms
        in Equation~\eqref{eq:latency}.
  \item Placement $A(t)$ fixes which inter-node latencies
        $L_{i,j}(t)$ each edge (dependent service pair) experiences in Equation~\eqref{eq:edge-latency-multi}.
  \item Call-graph demand, as summarized by $\{r_e(t)\}$ and
        $\{R_s^{\text{in}}(t)\}$, reshapes both CPU demand and the
        contribution of each edge (dependent service pair) to end-to-end latency.
\end{itemize}
These couplings make purely placement-only or scaling-only strategies suboptimal: a bad placement magnifies network delay even with sufficient replicas, while poor replica provisioning creates compute bottlenecks even under favorable network conditions.
Therefore, \textbf{\system treats \emph{placement} and \emph{replica provisioning} as a joint optimization problem driven by call-graph statistics, demand matrices, latency measurements, and resource constraints.}

\section{Problem Formulation}
\label{sec:formulation}

Let the measured and estimated system state at time $t$ be
\[
  S(t)
  \;=\;
  \big\{\,L(t),\;\{r_e(t),w_e(t),b_e(t)\}_{e\in E},\;
            \{R_s^{\text{in}}(t),W_s^{\text{in}}(t)\}_{s\in\mathcal{S}}\,\big\},
\]
where $L(t)=\{L_{i,j}(t)\}$ is the inter-node latency matrix,
$r_e(t),w_e(t),b_e(t)$ are edge-level demand statistics from
Section\ref{subsec:edge-stats}, and $R_s^{\text{in}}(t),W_s^{\text{in}}(t)$ are
service-level demands from Section\ref{subsec:service-demand}.

\subsection{Latency-weighted objective}
\label{subsec:objective}

Given state $S(t)$, \system jointly determines a replica allocation matrix
$A=[A_{s,n}]$ and the replica counts $x_s=\sum_{n}A_{s,n}$ that minimize total system cost while meeting service-level objectives.

We define a constant value for per-service resource cost with per-replica unit cost $c_s$ (e.g., normalized CPU cost) for running service $s$, and a network latency cost that weights edges by call rate:
\begin{equation}
\label{eq:obj}
\begin{aligned}
\min_{A}\quad \text{Cost}(A;S(t))
&\triangleq
\underbrace{%
%\sum_{s\in\mathcal{S}} c_s \sum_{n\in\mathcal{N}} A_{s,n}%
\sum_{s\in\mathcal{S}} c_s x_s%
}_{\mathclap{\text{resource cost}}}
\\
&\quad+
\underbrace{%
\lambda_L \sum_{e=(u,v)\in E} r_e(t)\,\bar L_e\bigl(A,L(t)\bigr)%
}_{\mathclap{\text{latency cost}}}
\end{aligned}
\end{equation}

\noindent\text{s.t.}\;
\begin{align}
\sum_{s\in\mathcal{S}} A_{s,n}\,R_s &\le C_n,
&& \forall\, n\in\mathcal{N}, \label{eq:cap}\\
Q_q\!\left[Y(x(A),A;S(t))\right] &\le \tau_{\text{e2e}}, \label{eq:sla-e2e}\\
Q_q\!\left[Y_s(x(A),A;S(t))\right] &\le \tau_s,
&& \forall\,s\in\mathcal{S}_{\text{SLO}}, \label{eq:sla-local}\\
A_{s,n} &\in \mathbb{N},\quad A_{s,n}\ge 0.
\end{align}

Here:
\begin{itemize}[leftmargin=*]
  \item $\lambda_L>0$ balances resource usage against network latency.
  \item $\bar L_e(A,L(t))$ is the replica-averaged edge latency defined in Equation~\eqref{eq:edge-latency-multi}.
  \item $Q_q[Y(\cdot)]$ denotes the $q$-quantile (e.g.,p50, p90, and p95) of the end-to-end latency distribution; $\tau_{\text{e2e}}$ is the configured end-to-end SLO for each types of root request.
  \item $Y_s(\cdot)$ denotes the per-service latency (from raw collected metrics spans), and $\tau_s$ are per-service p95 SLOs (from the SLO configuration file).
  In practice, $Y(x(A),A;S(t))$ and $Y_s(x(A),A;S(t))$ are not evaluated via the analytic model in Equation~\eqref{eq:latency}, but via recent latency samples collected from the tracing and metrics stack.

  \item $\mathcal{S}_{\text{SLO}}\subseteq\mathcal{S}$ is the subset of services with explicit local SLOs. 
\end{itemize}

Constraint Equation~\eqref{eq:cap} enforces per-node CPU and memory capacities. Constraint Equation~\eqref{eq:sla-e2e} requires the $q$-quantile of end-to-end latency to stay below the global SLO, while Equation~\eqref{eq:sla-local} enforces per-service SLOs where configured.

% AdaScale main framework
\begin{figure}[t]
    \centering
    \includegraphics[width=\linewidth]{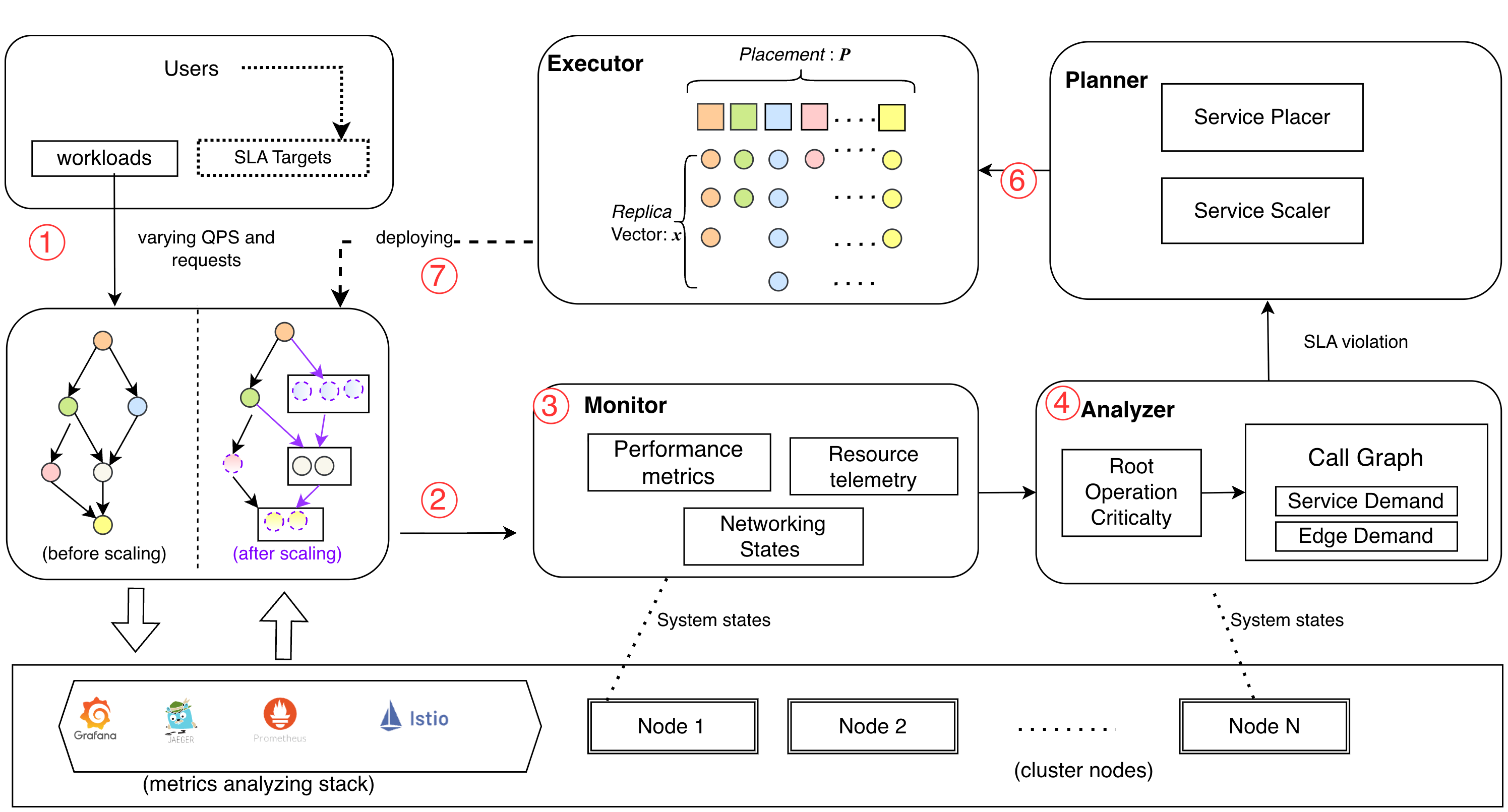}
    \caption{\system framework architecture and components.}
    \label{fig: AdaScale_framework}
\end{figure}

% two columns
% \begin{figure*}[!t]
%     \centering
%     \includegraphics[width=0.7\textwidth]{Figure/framework/AdaScale_v3.png}
%     \caption{\system framework architecture and components.}
%     \label{fig: AdaScale_framework}
% \end{figure*}

\subsection{Demand-driven capacity estimation}
Directly solving Equation~\eqref{eq:obj}--\eqref{eq:sla-local} in terms of $A_{s,n}$ is combinatorial and NP--hard. Instead, \system separates the calculation of \textbf{Scaling} (i.e., \emph{how many replicas}) from \textbf{Placement} (i.e., \emph{where to place them}).

Given service-level demand $\text{CPU\_demand}_s(t)$ from Section \ref{subsec:service-demand} and $\rho_s^{\max}\in(0,1)$  is the target max utilization per replica (e.g., $0.7$ for CPU), we first estimate the required replica count for service $s$ as
\begin{equation}
\label{eq:replicas-from-demand}
x_s^{\mathrm{dem}}(t)
\;=\;
\left\lceil
\frac{\text{CPU\_demand}_s(t)}{\mu_s\,\rho_s^{\max}}
\right\rceil,
\end{equation}

where $\mu_s$ is the estimated processing capacity
(CPU-seconds per second) of a single replica of $s$ at the full
utilization, \footnote{In practice, $\mu_s$ is calculated by recent Prometheus CPU usage and per-replica throughput metrics.} and per-replica safe capacity will be estimated by $\mu_s\,\rho_s^{\max}$.

This yields a demand-driven replica vector $x^{\mathrm{dem}}(t)=\{x_s^{\mathrm{dem}}(t)\}$, which is then rounded up to the closest integer and adjusted by the autoscaler to satisfy the SLO constraints.

Conditional on $x^{\mathrm{dem}}(t)$, the placement subproblem chooses a feasible $A$ with $\sum_n A_{s,n}=x_s^{\mathrm{dem}}$ that approximately minimizes the network latency term in Equation~\eqref{eq:obj}. This two-stage decomposition is realized by the two-timescale
adaptation mechanism via: (i) a slower demand-driven autoscaling loop that updates
$x^{\mathrm{dem}}(t)$; Compared with the current replica count $x^{\mathrm{current}}(t)$ the $x^{\mathrm{dem}}(t)$, the scaling actions of service $s$ at time $t$ will accordingly be chose from  \[
  ScaleActions(s,t)
  \;=\;
  \big\{\,Up(s,t), Down(s,t), Hold(s,t) \}
\]; and (ii) while keeping the replica count $x_s$ of service $s$ fixed,  a faster reactive placement loop that updates the replica placement matrix $A$.
In practice, the Analyzer enforces SLO constraints when determining replica targets, and the Planner assumes replica counts satisfy SLO feasibility and focuses on minimizing latency under placement constraints.

\subsection{Dynamic cross-node delays}
The latency matrix $L(t)$ is inherently time-varying due to congestion, background traffic, and externally injected dynamics (e.g., emulated delays for experiments). \system treats $L(t)$ as part of the observed state and reoptimizes placement whenever: 
(i) the change in average or high-percentile latency exceeds a threshold; or
(ii) the end-to-end SLO Equation~\eqref{eq:sla-e2e} is violated persistently.

Concretely, the reactive placement loop reduces an approximate
objective

% ingnore the migration cost here
\[
\min_{A'}\;
\sum_{e=(u,v)} r_e(t)\,\bar L_e\bigl(A',L(t)\bigr)
\,
\]
% \[
% \min_{A'}\;
% \sum_{e=(u,v)} r_e(t)\,\bar L_e\bigl(A',L(t)\bigr)
% \;+\;
% \beta\,\text{MigCost}(A,A'),
% \]
% where $\text{MigCost}(A,A')$ captures migration overheads and disruption risks, and $\beta\ge 0$ controls migration aggressiveness.

By weighting links by $r_e(t)$, the controller focuses migration budget on edges that both (i) are heavily exercised by the current workload and (ii) suffer from increased cross-node delays.

\section{Proposed AdaScale Framework}
\label{sec:architecture}

\system is a demand- and latency-aware microservice resource manager that
implements a Monitor--Analyzer--Planner--Executor (MAPE) control loop.
At a high level, \system repeatedly (i) monitors application and cluster state,
(ii) analyzes SLO risk and demand under the current workload mix, (iii) plans a
joint scaling and placement update, and (iv) executes the update safely using
native cluster mechanisms. A key design principle is a \emph{separation of
concerns}: the analysis logic determines \emph{how many} replicas are required
per service to satisfy SLOs, while the placement logic determines \emph{where}
to place those replicas to minimize latency under time-varying inter-node
delays.

%\paragraph{Decision epoch.}
At each control epoch $t$, \system produces a placement plan consisting of:
(i) desired replica totals $\{\hat x_s(t)\}_{s\in\mathcal{S}}$ and
(ii) a replica--node assignment matrix
$A^\star(t)=\bigl[A^\star_{s,n}(t)\bigr]$ (Section~\ref{sec:system-model}),
where $A^\star_{s,n}(t)$ is the number of replicas of service $s$ planned to run
on node $n$. The plan is computed from the measured state
$S(t)$ (Section~\ref{sec:formulation}), including demand estimates and the
inter-node latency matrix $L(t)$.

\vspace{0.25em}
\noindent The control loop mainly comprises five stages: %(Figure~\ref{fig: AdaScale_framework}):

\begin{itemize}[leftmargin=*]
  \item \textbf{Step~1 (Monitor: cluster exporters).}
        \system collects cluster state including node capacities,
        the current replica assignment matrix $A(t)$, and the inter-node latency
        matrix $L(t)$.

  \item \textbf{Step~2 (Monitor/Analyze: application telemetry and demand tables).}
        \system collects application telemetry (distributed traces and service
        metrics) and constructs demand summaries: per-edge demand statistics
        (call rates/work/bytes) and per-service demand statistics
        (aggregate call rates and CPU-demand proxies).

  \item \textbf{Step~3 (Analyzer: SLO- and demand-aware scaling decisions).}
        \system evaluates end-to-end and per-service SLO status, identifies the
        most critical root operations under a mixed workload, and derives
        per-service scaling actions (replica targets $\hat x_s(t)$) using a
        combination of trace-based criticality and demand estimates.

  \item \textbf{Step~4 (Planner: latency-aware placement).}
        Given replica targets and the current cluster network state $L(t)$,
        \system computes a new assignment matrix $A^\star(t)$ that reduces the
        demand-weighted latency objective while respecting node capacities.

  \item \textbf{Step~5 (Executor: actuation).}
        \system enforces $\{\hat x_s(t)\}$ and $A^\star(t)$ through safe rollouts
        and scheduling constraints, optionally with bounded parallelism
        $\kappa\in\mathbb{N}$ to reduce actuation time.
\end{itemize}

This pipeline aligns with the state definition $S(t)$ and the optimization goal
in Section~\ref{sec:formulation}. We next detail each component.

\subsection{Monitor}
\label{subsec:monitor}

% ALgo1 : Service Scaler
\begin{algorithm}[!t]
% \caption{Analyzer: multi-root criticality and SLO-/demand-aware scaling}
\caption{\textit{Service Scaler} with multi-root criticality and SLO-/demand-aware scaling}
\label{alg:analyzer}
\KwIn{
Root request operations $\mathcal{O}$ with $(\tau^{\mathrm{e2e}}_o,\pi_o,\mathcal{S}_o)$;
service SLOs $\{\tau_s\}$;
observed per-root p95 $\{\widehat{Y}^{95}_o\}$ from traces;
observed per-service p95 $\{\widehat{Y}^{95}_s\}$ and utilization $u_s$ from metrics;
%(optional) demand tables providing $\mathrm{CPU\_demand}_s$ and edge rates $r_{(u,v)}$
demand tables providing $\mathrm{CPU\_demand}_s$ and edge rates $r_{(u,v)}$
}
\KwOut{Replica targets $\{\hat{x}_s\}$ and weighted edges $\mathcal{E}(t)$}

\tcp{Root operation priority}
\ForEach{$o \in \mathcal{O}$}{
$v_o \leftarrow \max\{\widehat{Y}^{95}_o/\max(\tau^{\mathrm{e2e}}_o,1)-1,\,0\}$\;
$\kappa_o \leftarrow \pi_o \cdot v_o$\;
}
 Select target root op $o^\star$ (e.g., max criticality)\;

\tcp{Service priorities}
$\mathcal{S}^\star \leftarrow \bigcup \mathcal{S}_{o^\star}$ \tcp*{ Services triggered by $o^\star$.} 
\ForEach{$s \in \mathcal{S}^\star$}{
$\rho_s \leftarrow \widehat{Y}^{95}_s/\max(\tau_s,1)$ \tcp*{pressure ratio}
Compute trace-based criticality $\mathrm{crit}_s$ over $W_t$\;
Compute demand share $\mathrm{dem\_share}_s$ if demand tables exist (else set $\mathrm{dem\_share}_s \leftarrow 0$)\;
$\eta_s \leftarrow \alpha\cdot \mathrm{crit}_s + (1-\alpha)\cdot \mathrm{dem\_share}_s$\;
$\mathrm{score}_s \leftarrow \rho_s \cdot \eta_s$\;
Initialize $\hat{x}_s \leftarrow x^{\mathrm{cur}}_s$\;
}

\tcp{Propose actions with hysteresis}
\ForEach{$s \in \mathcal{S}^\star$}{
  % default action
  $\hat{x}_s \leftarrow x^{\mathrm{cur}}_s$; \tcp*{\texttt{hold} by default}
  \If{$\rho_s > \theta_{\uparrow}$ \textbf{or} $x^{\mathrm{dem}}_s > x^{\mathrm{cur}}_s$}{
    $\texttt{action}_s \leftarrow \mathtt{scale\_up}$\;
    $\hat{x}_s \leftarrow x^{\mathrm{cur}}_s + 1$\;
  }
  % \BlankLine
  \If{$\rho_s < \theta_{\downarrow}$ \textbf{and} $u_s < u_{\downarrow}$}{
    $\texttt{action}_s \leftarrow \mathtt{scale\_down}$\;
    $\hat{x}_s \leftarrow \max\!\big(x^{\mathrm{min}}_s,\; x^{\mathrm{cur}}_s - 1\big)$\;
  }
  \Else{
    $\texttt{action}_s \leftarrow \mathtt{hold}$\;
  }
}

\tcp{Scale-up budget}
Let $\mathcal{U} \leftarrow \{s\in\mathcal{S}^\star: \hat{x}_s > x^{\mathrm{cur}}_s\}$\;
Keep only the top-$K$ services in $\mathcal{U}$ by $\mathrm{score}_s$ as scale-up; set others to \texttt{hold} ($\hat{x}_s \leftarrow x^{\mathrm{cur}}_s$)\;

\tcp{Export edge weights for placement}
Construct $\mathcal{E}(t)=\{(u,v,\omega_{u,v})\}$ from demand/trace edges relevant to $\mathcal{O}^\star$\;
% Set $\omega_{u,v}\propto r_{(u,v)}$ (or fallback to per-trace edge frequency)\;
Set $\omega_{u,v}\propto r_{(u,v)}$\;
\Return $\{\hat{x}_s\},\mathcal{E}(t)$\;
\end{algorithm}

% Algo2: Service Placer
\begin{algorithm}[t]
% \caption{Planner: greedy latency- and capacity-aware replica placement}
\caption {\textit{Service Placer} with greedy latency- and capacity-aware replica placement}
\label{alg:planner}
\KwIn{
  Current placement $A$; target replicas $\{\hat{x}_s\}$;
  latency matrix $L(t)$; weighted edges $\mathcal{E}(t)$;
  node capacities and per-replica requests $\{R_s\}$
}
\KwOut{Planned placement $A^\star$}

$A^\star \leftarrow A$\;

\BlankLine
\Fn{\Cost{$A^\star$}}{
  $\mathrm{LatCost} \leftarrow \sum_{(u,v,\omega)\in\mathcal{E}(t)} \omega\cdot \bar{L}_{(u,v)}(A^\star,L(t))$\;
  % $\mathrm{CapPenalty} \leftarrow$ \text{large penalty for any node where requested CPU/mem exceeds capacity}\;
    $\mathrm{CapPenalty} \leftarrow$ \text{large penalty for exceeds capacity}\;
  \Return{$\mathrm{LatCost}+\mathrm{CapPenalty}$}\;
}

\BlankLine
\ForEach{service $s$}{
  \While{$\sum_n A^\star_{s,n} < \hat{x}_s$}{
    Find node $n^\star \in \mathcal{N}$ minimizing \textsc{Cost}($A^\star + \mathbf{1}_{(s,n)}$)\;
    $A^\star_{s,n^\star} \leftarrow A^\star_{s,n^\star}+1$\;
  }
  \While{$\sum_n A^\star_{s,n} > \hat{x}_s$}{
    Let $\mathcal{N}_s=\{n: A^\star_{s,n}>0\}$\;
    Find node $n^\star \in \mathcal{N}_s$ minimizing \textsc{Cost}($A^\star - \mathbf{1}_{(s,n)}$)\;
    $A^\star_{s,n^\star} \leftarrow A^\star_{s,n^\star}-1$\;
  }
}
\Return{$A^\star$}\;
\end{algorithm}

The Monitor produces a time-indexed snapshot of the system state used by
subsequent stages. Conceptually, it maintains a cache $\mathcal{D}(t)$, which includes 
current replica placement matrix $A(t)$, current inter-node latency matrix $ L(t)$, \text{node capacities}, \text{service metrics}, and \text{demand tables}.
% % \[
% % \mathcal{D}(t)
% % =
% % \Bigl\{
% % A(t),\; L(t),\; \text{node capacities},\; \text{service metrics},\;
% % \text{demand tables}
% % \Bigr\},
% % \]
% where $A(t)$ is the current replica placement matrix and $L(t)$ is the current
% inter-node latency matrix.

\subsubsection{Workload and performance telemetry}
\system collects (i) distributed traces to recover call-graph structure and
end-to-end latency samples, (ii) per-service performance metrics such as
p50/p90/p95 latencies and error rates, and (iii) resource telemetry such as CPU
and memory usage. These measurements are aggregated over a short window $W_t$ to
provide a stable estimate of the current operating regime.

\subsubsection{Demand tables (call-graph demand estimator)}
To make scaling and placement explicitly \emph{demand-aware}, \system computes
two demand summaries from the same telemetry window:
\begin{itemize}[leftmargin=*]
  \item an \emph{edge-demand} table that estimates, for each directed service
        dependency $e=(u,v)$, its call rate $r_e(t)$ (calls/s), expected work
        $w_e(t)$ (ms, derived from span durations), and (when available) byte
        rate $b_e(t)$ (bytes/s);
  \item a \emph{service-demand} table that aggregates incoming/outgoing rates
        and work, yielding $R_s^{\text{in}}(t)$, $R_s^{\text{out}}(t)$,
        $W_s^{\text{in}}(t)$, and a CPU-demand proxy
        $\mathrm{CPU\_demand}_s(t)\approx R_s^{\text{in}}(t)\cdot W_s^{\text{in}}(t)/1000$
        (Section~\ref{subsec:service-demand}).
\end{itemize}
These demand tables instantiate the quantities in our system model and are
reused by both the scaling logic (Analyzer) and the placement logic (Planner).
%\textcolor{red}{(Component: Call-Graph Demand Estimator)}

\subsubsection{Cluster resource and placement telemetry}
\system snapshots node capacities (CPU and memory) and the current replica
assignment matrix $A(t)$. For each service $s$, $A_{s,n}(t)$ records how many
replicas are currently placed on node $n$, and $x_s(t)=\sum_n A_{s,n}(t)$ is the
service replica count. The snapshot also records per-replica resource requests
$R_s$ needed for capacity checks in the planner.

\subsubsection{Networking state (inter-node latency)}
\system continuously measures or infers a matrix of inter-node latencies
$L(t)=\{L_{i,j}(t)\}$.
We treat $L(t)$ as a first-class runtime signal: changes in $L(t)$ can trigger
placement updates even if the workload mix is unchanged. This design captures
dynamic environments (e.g., congestion, background interference, or injected
delays) without requiring explicit bandwidth modeling.

\subsection{Analyzer}
\label{subsec:ingest}

The Analyzer transforms $\mathcal{D}(t)$ into per-service scaling actions and
a compact, demand-weighted call-graph representation for placement. The Analyzer
implements two coupled tasks: (i) \emph{SLO risk assessment under mixed root
operations} and (ii) \emph{demand-aware scaling decisions}.
%\textcolor{red}{(Component: Ingest/Autoscaling Module)}

\subsubsection{Root-operation criticality under mixed workloads}
Microservice applications often serve multiple root request classes
$\mathcal{O}=\{o_1,\ldots,o_K\}$, each with its own end-to-end SLO target
$\tau^{\mathrm{e2e}}_o$ and workload proportion $\pi_o$ (from configuration or
telemetry, with $\sum_o \pi_o=1$). The Analyzer computes an observed end-to-end
p95 latency $\widehat{Y}_o^{(95)}(t)$ per root operation and defines a
workload-weighted SLO risk score:
\[
v_o(t)\triangleq \max\left\{\frac{\widehat{Y}_o^{(95)}(t)}{\max\{\tau^{\mathrm{e2e}}_o,1\}}-1,\;0\right\},
\qquad
\kappa_o(t)\triangleq \pi_o\cdot v_o(t).
\]
The system prioritizes root operations with large $\kappa_o(t)$, ensuring that
optimization effort is focused on request classes that are both frequent and
SLO-threatening.

\subsubsection{Service criticality and scaling actions}
To select which services to scale, \system computes a trace-based service
criticality score $\mathrm{crit}_s(t)$,
and combines it with demand signals such as $\mathrm{CPU\_demand}_s(t)$ or its
normalized share. Given per-service SLO thresholds $\tau_s$ and observed p95
latencies $\widehat{Y}_s^{(95)}(t)$, the Analyzer derives a pressure ratio
$\rho_s(t)=\widehat{Y}_s^{(95)}(t)/\max\{\tau_s,1\}$ and proposes an action
$a_s(t)\in\{\mathrm{scale\_up},\mathrm{scale\_down},\mathrm{hold}\}$ together
with a replica target $\hat x_s(t)$.
To avoid oscillations and limit control-plane disruption, \system enforces a
budget: only the top-$K$ services by a composite priority (pressure combined
with criticality/demand) are allowed to scale up in each epoch, while remaining
services are held.

\subsubsection{Demand-weighted edge set for placement}
For placement planning, the Analyzer exports a weighted edge set
$\mathcal{E}(t)=\{(u,v,\omega_{u,v}(t))\}$, where $(u,v)$ is a service dependency
and $\omega_{u,v}(t)$ is proportional to the edge demand (e.g., call rate).
This edge set summarizes which dependencies dominate end-to-end performance
under the current workload mix and provides the primary signal for
latency-aware placement.

\subsection{Planner}
\label{subsec:controller}

The Planner computes a replica assignment plan $A^\star(t)$ given
(i) replica targets $\{\hat x_s(t)\}$ from the Analyzer, (ii) current placement
$A(t)$, (iii) node capacities, and (iv) the measured inter-node latency matrix
$L(t)$. 
%\textcolor{red}{(Component: Placement Controller)}

\subsubsection{Latency- and capacity-aware objective}
The Planner minimizes a demand-weighted latency objective subject to capacity
constraints. Using the weighted edge set $\mathcal{E}(t)$ and the replica-averaged
edge latency $\bar L_{(u,v)}(A,L(t))$ (Equation~\eqref{eq:edge-latency-multi}), it
evaluates:
\[
\mathrm{LatCost}(A)
=
\sum_{(u,v)\in\mathcal{E}(t)}
\omega_{u,v}(t)\cdot \bar L_{(u,v)}(A,L(t)).
\]
To discourage capacity violations, it adds a large penalty when the implied CPU
or memory requests exceed any node capacity, yielding
$\mathrm{Cost}(A;S(t))=\mathrm{LatCost}(A)+\mathrm{CapPenalty}(A)$ as described in
Section~\ref{subsec:controller}.

\subsubsection{Greedy replica placement and removal}
As the joint problem is combinatorial, the Planner uses a greedy local
search guided by $\mathrm{Cost}(A;S(t))$. For each service $s$, it compares the
current replica count $x_s(t)=\sum_n A_{s,n}(t)$ with the target $\hat x_s(t)$:
if $\hat x_s(t)>x_s(t)$ it adds replicas one-by-one to the node that yields the
lowest cost; if $\hat x_s(t)<x_s(t)$ it removes replicas one-by-one from the
node whose removal causes the smallest cost increase. This produces a new
assignment $A^\star(t)$ that is simultaneously demand-aware (through
$\omega_{u,v}(t)$), latency-aware (through $L(t)$), and capacity-aware (through
resource penalties).

\subsection{Executor}
\label{subsec:executor}

The Executor actuates the scaling and placement decisions produced by the
Planner. Its input is the plan $(\{\hat x_s(t)\},A^\star(t))$ for the current
epoch. For each service $s$, it (i) enforces the replica target $\hat x_s(t)$
and (ii) constrains scheduling to the allowed node set
$\mathcal{N}_s=\{n\mid A^\star_{s,n}(t)>0\}$ so that replicas run on planner-chosen
nodes. When feasible, it additionally applies spreading rules that bias replicas
toward the intended distribution encoded by $A^\star(t)$. The Executor monitors
rollout progress and declares success once the cluster converges to a state
consistent with the plan (subject to timeouts). To reduce actuation latency, it
supports bounded parallelism with a configurable concurrency limit $\kappa$.
%\textcolor{red}{(Component: Executor)}

% \paragraph{End-to-end workflow.}
When putting all stages together, each decision epoch:
(i) refreshes $\mathcal{D}(t)$ from telemetry,
(ii) computes scaling actions and edge weights in the Analyzer,
(iii) computes $A^\star(t)$ in the Planner under the current $L(t)$, and
(iv) enforces the resulting plan in the Executor. This completes one MAPE
iteration and realizes the two-stage decomposition: \emph{how many} replicas are needed is decided by SLO- and demand-driven analysis, while \emph{where} to place replicas is
decided by latency-aware planning under dynamic cross-node delays.

\section{Performance Evaluation}
\label{sec:performance evaluation}

\subsection{Cluster Testbed}
We prototype and evaluate \system on a Kubernetes-based cloud--edge testbed~\cite{kubernetes}. The testbed comprises 16 virtual machines (VMs): one control-plane node and fifteen worker nodes. The control-plane VM is equipped with 32 CPU cores (AMD x86\_64), 32\,GiB RAM, and a 16\,Gbps network interface. Each worker VM provides 8 CPU cores from the same EPYC processor family, 32\,GiB RAM, and a 16\,Gbps link.

\textbf{Software stack.}
All nodes run Ubuntu 22.04.2 LTS with Linux kernel 5.15.0. We deploy Kubernetes v1.27.4 on ten nodes and Kubernetes v1.28.4 on the remaining five nodes. Calico v3.26.1 is used as the CNI, Istio v1.20.3 provides the service mesh, and CRI-O serves as the container runtime (v1.27.1 on ten nodes and v1.28.11 on five nodes).

\textbf{Intra-cluster connectivity and network dynamics.}
The VMs are hosted in the university’s dedicated research cloud. As a result, the baseline inter-node round-trip latency is very small (typically 0.2--1\,ms), which we confirmed using ICMP ping measurements. To emulate cloud--edge networking dynamics in a controlled and repeatable manner, \system programmatically applies cross-node delay constraints within the cluster, inspired by TraDE~\cite{TraDE_TPDS2026}. Unless stated otherwise, we refresh these injected conditions periodically during experiments so that we can assess how well the evaluated policies adapt to time-varying network environments.

\textbf{Benchmark application and request generation.}
We use the Social Network application from DeathStarBench~\cite{deathStarBench_ASPLOS19} as the representative microservice workload, and we generate client requests using \texttt{wrk2}~\cite{wrk2}. Social Network implements a simplified social-media service and consists of 27 microservices supporting operations such as composing posts as well as reading user and home timelines. Different request types exercise different call-graph structures and traffic patterns, enabling us to evaluate \system under diverse end-to-end execution paths.

% For all policy evaluations, we focus on end-to-end response time for the front-end service (\texttt{nginx-thrift}) and report three metrics: average latency (Avg), 99th percentile latency (p99), and standard deviation (Stdev). We define SLA thresholds on average response time that depend on cluster scale: 100 $ms$ for five nodes, 150 $ms$ for ten nodes, and 200 $ms$ for fifteen nodes, matching the thresholds annotated in Table~\ref{tab:policy1_evaluation}. For the hybrid-dynamics experiments, a less stringent SLA threshold of 300 $ms$ is used to highlight how policies behave under injected high-latency phases.

%15 cluster-node
% hardware, software

% Table 1: Per-edge demand statistics.

\subsection{Root Requests Analysis}
In the Social Network benchmark, there are three types of root request operations. To define the SLO targets for each root request, different QPS are used to measure end-to-end performance and identify the kneepoint at which it is significantly affected. When QPS (Queries Per Second) increases, the tail latency of the corresponding microservice first gradually increases, then suddenly increases sharply at a specific QPS and latency value. We refer to such a point as a \textit{kneepoint} to represent the QoS-violation point for a specific root operation request. Thus, under our experiment environment, we define the SLO target of a specific root operation by \textit{kneepoint QPS}, where \textit{max\_load} QPS is reached.

\begin{figure*}[htbp]
  \centering
  % ---------- (a) ----------
  \subfloat[Kneepoint QPS exploration of root request $\lambda_1(t)$\label{fig:knee_request1}]{
    \includegraphics[width=.32\linewidth]{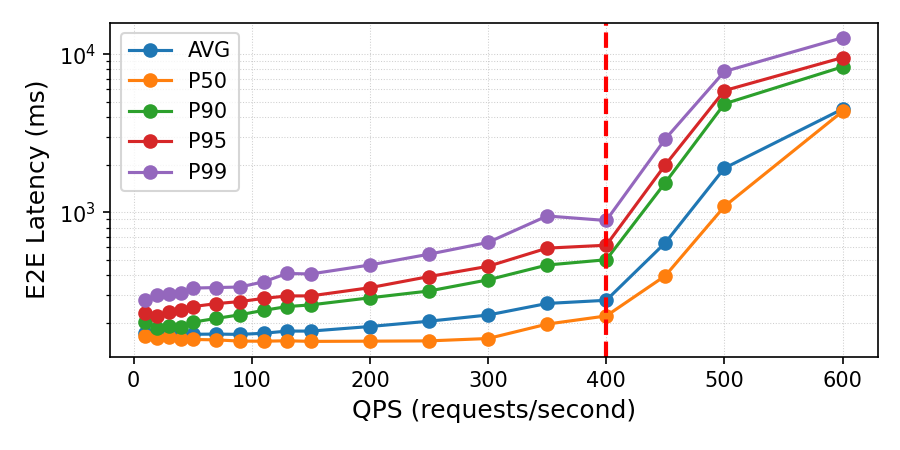}}
  \hfill
  % ---------- (b) ----------
  \subfloat[Kneepoint QPS exploration of root request $\lambda_2(t)$\label{fig:knee_request2}]{
    \includegraphics[width=.32\linewidth]{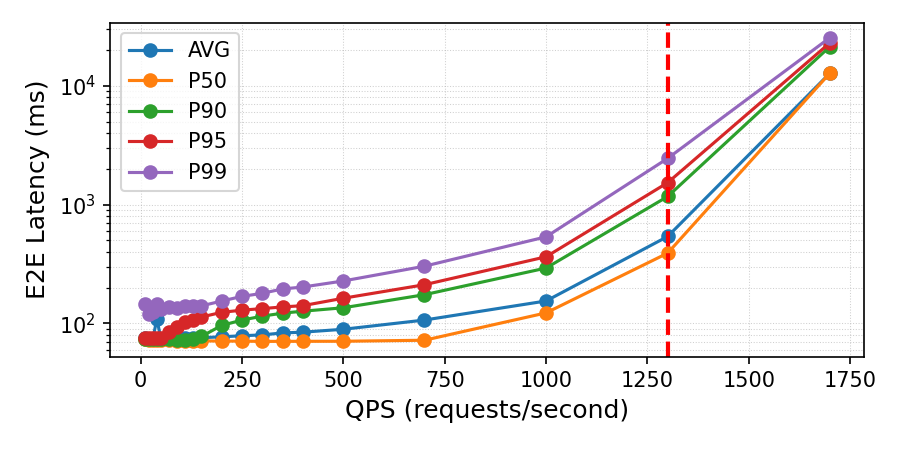}}
  \hfill
  % ---------- (c) ----------
  \subfloat[Kneepoint QPS exploration of root request $\lambda_3(t)$\label{fig:knee_request1}]{
    \includegraphics[width=.32\linewidth]{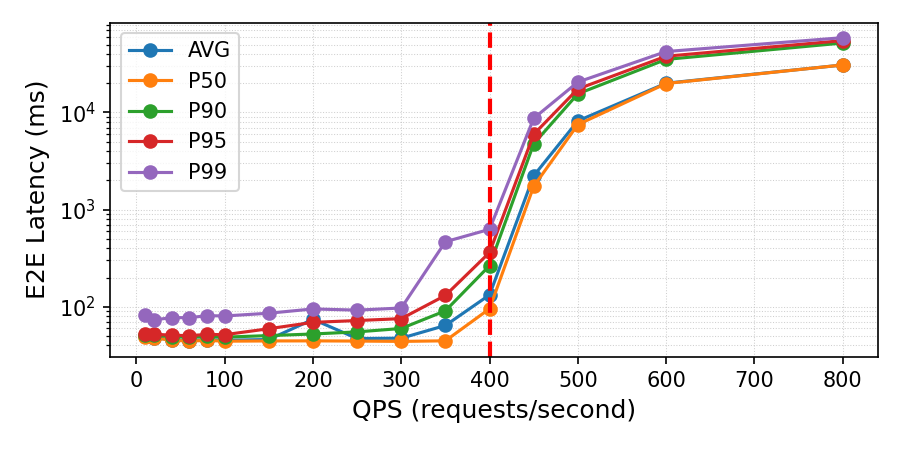}}

  \caption{In the Social Network benchmark, there are three types of root requests. To define the SLO for each root request, different QPS are sent to explore the kneepoints, at which the end-to-end performance is significantly affected.}
  \label{fig:social_net_root_ops}
\end{figure*}

To find the \textit{kneepoint QPS} of each root request operation of the deployed social network application, we conducted load tests for each root request operation by gradually increasing the corresponding QPS from lower values to higher values, which saturate the deployed microservice application under each root request operation. As shown in Figure \ref{fig:social_net_root_ops}, we demonstrated different percentile metrics including AVG (average), p50 (median), p90 (90th Percentile), p95 (95th Percentile), and p99 (99th Percentile) for the response time under each root request. For all the root request operations, i.e., $\lambda_1(t)$, $\lambda_2(t)$ and $\lambda_3(t)$ , each percentile metric shows similar increasing trends as the QPS increases. Specifically, the kneepoint QPS for each root operation is approximately 400, 1300, and 400.

\begin{table}[t]
  \centering
  \caption{Per-edge demand statistics.}
  \label{tab:edge-demand}
  \begin{tabular}{ccccccc}
    \toprule [1.2pt]
    \textbf{Src} & \textbf{Dst} & $p_e$ & $r_e^{\text{per-req}}$ & $w_e$ [ms] & $r_e$ [calls/s] & $b_e$ [bytes/s] \\
    \midrule\hline
    S0 & S1  & 0.106 & 1.063 & 275.376 & 10.63  & --    \\
    S0 & S2  & 0.311 & 3.110 & 3.884   & 31.102 & --    \\
    S0 & S3  & 0.567 & 5.669 & 2.618   & 56.693 & --    \\
    S1 & S2  & 0.106 & 1.063 & 2.401   & 10.63  & 8913  \\
    S1 & S3  & 0.106 & 1.063 & 40.996  & 10.63  & 9281  \\
    S1 & S4  & 0.106 & 1.063 & 87.745  & 10.63  & 73395 \\
    S1 & S5  & 0.106 & 1.063 & 0.023   & 10.63  & 8237  \\
    S1 & S6  & 0.106 & 1.063 & 0.012   & 10.63  & 10886 \\
    S1 & S7  & 0.106 & 1.063 & 0.014   & 10.63  & 13201 \\
    S1 & S8  & 0.106 & 1.063 & 10.161  & 10.63  & 52095 \\
    S2 & S8  & 0.311 & 3.110 & 0.011   & 31.102 & 21564 \\
    S3 & S8  & 0.559 & 5.591 & 0.011   & 55.906 & 43380 \\
    S3 & S9  & 0.106 & 1.063 & 38.724  & 10.63  & 7808  \\
    S4 & S10 & 0.106 & 1.063 & 24.360  & 10.63  & 10915 \\
    S4 & S11 & 0.106 & 1.063 & 1.497   & 10.63  & 39275 \\
    \bottomrule[1.2pt]
  \end{tabular}
\end{table}

% kneepoints QPS, proportions

\subsection{Service and Edge Demands in Call Graphs}
To get fine-grained statistics of the call graph at run time, we define the service demand and edge demand in Section~\ref{subsec:service-demand}.
In the evaluation, we present a snapshot of the runtime service and edge demand under a 5-minute sustained mix workload (root requests proportion 1:3:6). Table ~\ref{tab:edge-demand} shows per-edge demand statistics inferred from Jaeger traces and Prometheus traffic for a single time window $W_t$. For each edge $e=(u\!\to\!v)$ we list the empirical probability $p_e$, the expected number of traversals per root request $r_e^{\text{per-req}}$, the mean edge-local service time $w_e$ in milliseconds, the estimated call rate $r_e$ (calls/s), and the byte rate $b_e$ (bytes/s).

Under 5 minutes sustained mix workload (root requests proportion $\lambda_1: \lambda_2: \lambda_3 =1:3:6$) to the deployed microservice application, per-service demand statistics aggregated from edge-level demand in Table~\ref{tab:edge-demand} for the same time window $W_t$. For each service $s$, we list $R_s^{\text{in}}$, $R_s^{\text{out}}$, $W_s^{\text{in}}$, and $B_s^{\text{in}}$ as defined in Section~\ref{subsec:service-demand}. Besides, Table~\ref{tab:service-demand} demonstrates the per-service demand statistics aggregated from edge-level demand in Table~\ref{tab:edge-demand} for the same time window $W_t$. For each service $s$, we list $R_s^{\text{in}}$, $R_s^{\text{out}}$, $W_s^{\text{in}}$, and $B_s^{\text{in}}$ as defined in Section~\ref{subsec:service-demand}. 

In Table~\ref{tab:edge-demand}, $S0$ is the entry service which receives root requests and triggers the downstream services $S1$, $S2$ and $S3$. It can be observed that the probability $p_e$ of the root requests exercising the edges ($S0\rightarrow S1$, $S0\rightarrow S2$, $S0\rightarrow S3$) are 0.106, 0.311, and 0.567, which are almost the same as the proportions of each request ($\lambda_1: \lambda_2: \lambda_3 =1:3:6$). For Table ~\ref{tab:service-demand}, the in-degree and out-degree counts of each service exactly match the actual service call graph structure in Social Network \cite{deathStarBench_ASPLOS19}. For instance, considering the entry service $S0$, the in-degree count is 0 because $S0$ is the entry service of the call graph, thus there is no upstream service for $S0$, and the out-degree count is 3 because the downstream services of $S0$ consist of $S1$, $S2$, and $S3$. \textbf{Therefore, both tables accurately approximate the fine-grained statistics of the call graph}.

% Table2: Per-service demand statistics.
\begin{table}[t]
\centering
\caption{Per-service demand statistics.}
\label{tab:service-demand}
% \scriptsize % Reduce font size
\begin{tabularx}{\columnwidth}{l *{6}{>{\centering\arraybackslash}X}}
\toprule [1.2pt]
\textbf{Service} & in\_deg & out\_deg & $R_s^{\text{in}}$ & $R_s^{\text{out}}$ & $W_s^{\text{in}}$ & $B_s^{\text{in}}$ \\ 
 &  &  & [calls/s] & [calls/s] & [ms] & [bytes/s] \\
\midrule \hline
S0  & 0 & 3 & 0      & 98.425  & 0       & 0      \\
S1  & 1 & 7 & 10.63  & 74.409  & 275.376 & 0      \\
S2  & 2 & 1 & 41.732 & 31.102  & 3.506   & 8913   \\
S3  & 2 & 2 & 67.323 & 66.535  & 8.678   & 9281   \\
S4  & 1 & 2 & 10.63  & 21.26   & 87.745  & 73395  \\
S5  & 1 & 0 & 10.63  & 0       & 0.023   & 8237   \\
S6  & 1 & 0 & 10.63  & 0       & 0.012   & 10886  \\
S7  & 1 & 0 & 10.63  & 0       & 0.014   & 13201  \\
S8  & 3 & 0 & 97.638 & 0       & 1.116   & 117039 \\
S9  & 1 & 0 & 10.63  & 0       & 38.724  & 7808   \\
S10 & 1 & 0 & 10.63  & 0       & 24.360  & 10915  \\
S11 & 1 & 0 & 10.63  & 0       & 1.497   & 39275  \\
\bottomrule [1.2pt]
\end{tabularx}
\end{table}

% service_aliases:
%   nginx-web-server: S0
%   compose-post-service: S1
%   user-timeline-service: S2
%   home-timeline-service: S3
%   text-service: S4
%   unique-id-service: S5
%   user-service: S6
%   media-service: S7
%   post-storage-service: S8
%   social-graph-service: S9
%   user-mention-service: S10
%   url-shorten-service: S11
\subsection{Performance Comparison}
% throughput 
% and normalized response-time SLA targets for each request
\subsubsection{Compared Methods}
To evaluate our proposed \system framework, we compared it with the  Kubernetes Horizontal Pod Autoscaler (HPA) policy and the traffic-aware NetMARKS\_Scale, which was adapted from \cite{NetMARKS_InfoCOM21}, targeting network-aware management for microservice applications. For K8s HPA, this is mainly achieved control loop that automatically adjusts the number of pod replicas in a workload (such as a Deployment or StatefulSet) to match observed demand. This is done to maintain performance and optimize resource usage without manual intervention. For the method of NetMARKS\_Scale, it adopts a similar service mesh with AdaScale, but focuses on optimizing the most communicated microservice pairs. 

% Normalized response time for Social Network
\begin{figure*}[htbp]
  \centering
  % ---------- (a) ----------
  \subfloat[root operation 1 \texttt{Compose Post}\label{fig:root1_RT}]{
    \includegraphics[width=.32\linewidth]{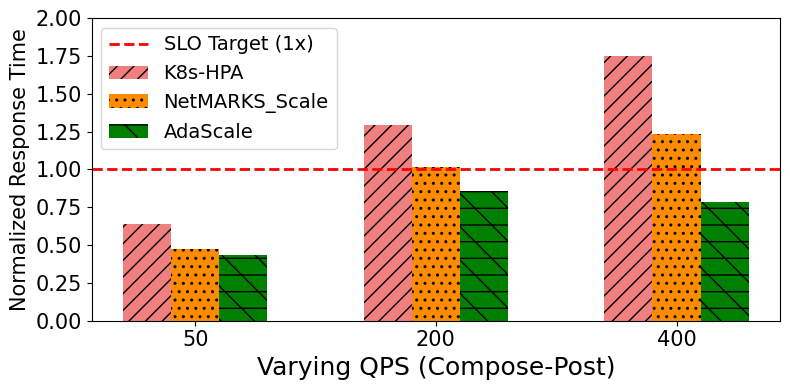}}
  \hfill
  % ---------- (b) ----------
  \subfloat[root operation 2 \texttt{Read User Timline}\label{fig:root2_RT}]{
    \includegraphics[width=.32\linewidth]{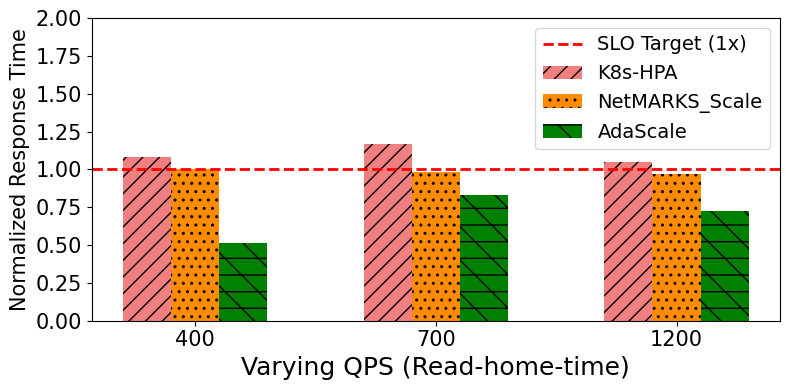}}
  \hfill
  % ---------- (c) ----------
  \subfloat[root operation 3 \texttt{Read Home timeline}\label{fig:root3_RT}\label{fig:root3_RT}]{
    \includegraphics[width=.32\linewidth]{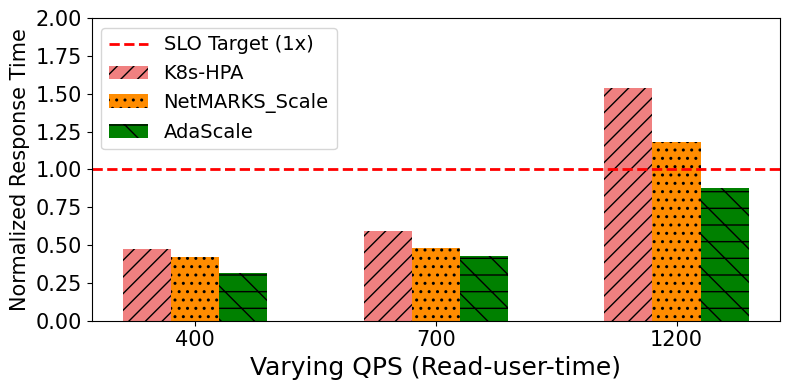}}

  \caption{Under each root requests and varying QPS, \system outperformed existing methods in terms of response time.}
  \label{fig:social_net_root_RT}
\end{figure*}

\subsubsection{End-to-end Performance}
The experiment results showed improved throughput and decreased average response time for different root operations (requests).

\textbf{Response Time.}
For the social network benchmark application, we used the wrk2 tool to generate three request types: \texttt{compose-post}, \texttt{read-user-timeline}, and \texttt{read-home-timeline}. Each type shows distinct call graphs and traffic patterns across the application’s dependency graph. Figure \ref{fig:social_net_root_RT} compares the normalized average response time under various workloads, including changes in QPS and root request operations. For these request types, \system consistently outperforms existing methods, meeting SLO targets across scenarios. Compared to NetMARKS\_Scale \cite{NetMARKS_InfoCOM21}, \system achieves up to 1.56\texttt{x} lower response times for \texttt{compose-post}, 1.93\texttt{x} for \texttt{read-home-timeline}, and 1.34\texttt{x} for \texttt{read-user-timeline} requests. Thus, \system outperforms existing methods in terms of response time.

% Througput for Social Network
\begin{figure*}[htbp]
  \centering
  % ---------- (a) ----------
  \subfloat[root operation 1 \texttt{Compose Post}\label{fig:root1_THput}]{
    \includegraphics[width=.32\linewidth]{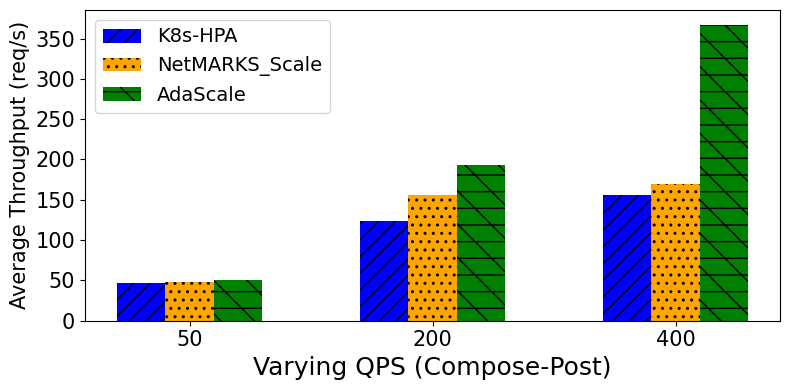}}
  \hfill
  % ---------- (b) ----------
  \subfloat[root operation 2 \texttt{Read User Timeline}\label{fig:root2_THput}]{
    \includegraphics[width=.32\linewidth]{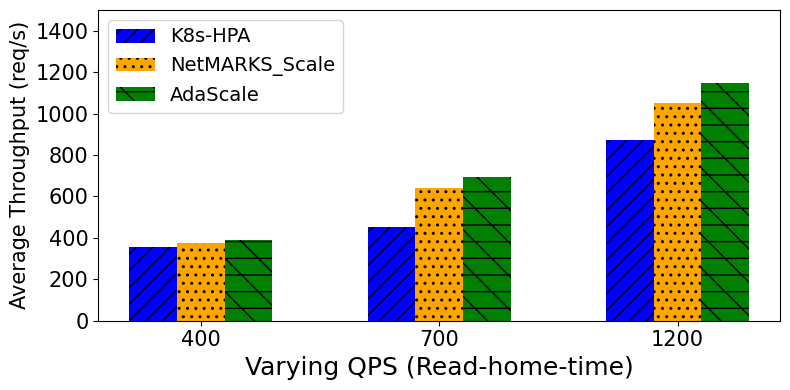}}
  \hfill
  % ---------- (c) ----------
  \subfloat[root operation 3 \texttt{Read Home timeline}\label{fig:root3_RT}\label{fig:root3_THput}]{
    \includegraphics[width=.32\linewidth]{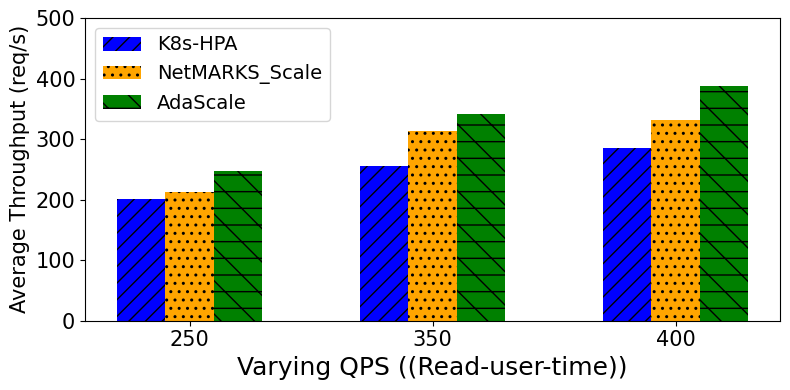}}

  \caption{Under three types of root request operations and varying QPS, \system outperformed existing methods in throughput.}
  \label{fig:social_net_root_THput}
\end{figure*}
\textbf{Throughput.} 
Throughput critical metrics for evaluating the deployed applications' end-to-end performance in dynamic computing environments. Throughput refers to the total quantity of transmitted data, including protocol overhead and retransmissions. In terms of \textit{throughput}, we conducted experiments to measure the average throughput (requests/second) with different QPS of workloads and mixed root request operations. As shown in the three subplots of Figure \ref{fig:social_net_root_THput}, the proposed \system achieves higher throughput across various scenarios. Specifically, compared with NetMARKS\_Scale \cite{NetMARKS_InfoCOM21}, our method shows an overall higher throughput and outperforms it by up to 2.16x for \texttt{compose-post} requests, 1.32x for \texttt{read-home-timeline} requests, 1.36x for \texttt{read-user-timeline} requests. Thus, compared with the existing methods, our proposed method \system achieves higher overall throughput across various scenarios.

\subsection{Limitations}
We have limitations in the diversity of multiple deployed microservice applications and in dynamic bandwidth conditions. We plan to extend the proposed \system in future work regarding the limitations.
% more test scenarios

\section{Related Work}
\label{sec:realted work}
% The management of microservices in different scenarios has been extensively explored. This section reviews existing literature emphasizing microservice management, call-graph analysis, and network-aware scheduling. 
% Besides, Table \ref{tab:table_relatedWork} compares \texttt{iDynamics} with representative work from different aspects.
%The dynamic management of microservices in cluster environments has been studied from various perspectives. This section reviews the existing literature, focusing on microservice management, call-graph dynamics analysis, network dynamics emulation, and network-aware scheduling.

\subsection{Microservice Management}
 FIRM~\cite{FIRM_MS}, Erms~\cite{Erms_ASPLOS23}, GrandSLAm~\cite{grandslam_MS}, and Sage~\cite{Sage_ASPLOS21} leverage online telemetry and modeling to localize SLO violations, assign latency budgets to services, and trigger fine-grained reprovisioning or request reordering in shared microservice environments. TraDE~\cite{TraDE_TPDS2026} is proposed to redeploy microservice instances to maintain QoS under changing workloads and network conditions.  CoScal \cite{ms_rl_TNSM} is presented as a multi-faceted scaling method integrating workload prediction methods and reinforcement learning. Yi Hu et al. investigate the dynamic service mesh orchestration by using probabilistic routing for microservice call graphs. Kai Peng et al. study large-scale service-mesh orchestration using Jackson queuing network theory and a three-stage heuristic \cite{collaborativeMesh_Infocom25}. However, these methods tend to ignore the influences of different types of root requests.

 % These research works demonstrate complex management logic but treat their execution environment largely as given: they do not expose a generic evaluation framework that allows different scheduling policies to be plugged in and compared under configurable call-graph, traffic, and networking dynamics. In contrast, \texttt{iDynamics} does not propose yet another resource manager; instead, it provides the infrastructure required to test such managers under realistic and repeatable cloud--edge dynamics.

% Delete to reduce page
\subsection{Call-Graph Dynamics Analysis}
Several works analyze microservice dependencies and their impact on performance. 
Tian et al.~\cite{TaskDependenceis_SoCC19} synthesize task-dependency graphs for data-parallel jobs from large-scale cluster traces. Yi Hu et al. introduce a joint optimization method for service deployment and request routing via fine-grained queuing network analysis \cite{deployment_routing_TPDS23}. Luo et al.~\cite{SoCC2021_MS_luo} characterize microservice call graphs from Alibaba traces, categorizing microservice dependencies into three distinct types. Parslo~\cite{Parslo_SoCC21} decomposes end-to-end SLO budgets into node-specific latency targets using gradient descent. Sage~\cite{Sage_ASPLOS21} also models dependencies when diagnosing QoS violations. However, these methods are generally time-consuming with high overheads to build the graph and are not suitable for dynamic incoming user requests.

\subsection{Network-aware Microservice Scheduling}
Network-aware microservice management methods exploit infrastructure-level metrics when placing microservice instances. Fangyu Zhang et al introduce a network-aware reliability model and optimized microservice placement algorithms to enhance service reliability and reduce bandwidth consumption in mobile and IoT networks \cite{Net_ms_placement_TNSM}. NetMARKS~\cite{NetMARKS_InfoCOM21} is proposed as a dynamic Kubernetes scheduling approach leveraging Istio~\cite{istio} network metrics to optimize containerized workflows for 5G edge applications. Marchese et al. extend the default Kubernetes scheduler with network-aware placement policies~\cite{ExtendingNetK8s_CCGRID22, ExtendingNetK8s_ICSOC22}. OptTraffic~\cite{OptTraffic_ICPP23} is a network-aware scheduling system that minimizes cross-machine traffic in containerized microservices under multi-replica deployment. However, these research works have limitations in considering the dynamic networking states across the cluster nodes.

% \subsection {workload dyanmics}
% \subsection {SLAs dyanmics}

\section{Conclusions and Future Work}
\label{sec:conclusion}
We proposed, designed and implemented \system, an adaptive scaling and placement framework for microservices under dynamics. Our framework mainly considers the dynamics, including cross-node delays, dynamic call graphs, and the root request operations. We designed \system as an MAPE (Monitor--Analyzer--Planner--Executor) control loop with different functionalities for each process. Besides, we proposed two algorithms for the service scaling and placement process with the metrics and defined statistics from the service call graph.
Experiments on a social network benchmark demonstrated the accuracy of root request operation modeling and showed that \system could outperform existing methods in response time and throughput under dynamic conditions.

\bibliographystyle{IEEEtran}
\bibliography{references}

\end{document}